\documentclass[12pt,onecolumn]{emulateapj}
\usepackage{natbib}

\newcommand{\pacz}{Paczy\'nski}

\shorttitle{M31 pixel lensing event OAB-N2}
\shortauthors{Calchi Novati et al.}

\begin{document}

\title{M31 pixel lensing event OAB-N2: 
a study of the lens proper motion}

\author{S.~Calchi Novati\altaffilmark{1,2,3},
M.~Dall'Ora\altaffilmark{4},
A.~Gould\altaffilmark{5},
V.~Bozza\altaffilmark{1,2,3},
I.~Bruni\altaffilmark{6},
F.~De Paolis\altaffilmark{7},
M.~Dominik\altaffilmark{8}\footnote{Royal Society University Research Fellow},
R.~Gualandi\altaffilmark{6},
G.~Ingrosso\altaffilmark{7}, 
Ph.~Jetzer\altaffilmark{9},
L.~Mancini\altaffilmark{1,2,3,10}, 
A.~Nucita\altaffilmark{11},
G.~Scarpetta\altaffilmark{1,2,3},
M.~Sereno\altaffilmark{12,13}, 
F.~Strafella\altaffilmark{7}
(PLAN collaboration\footnote{http://plan.physics.unisa.it/index.htm})}

\affil{$^1$ Dipartimento di Fisica ``E. R. Caianiello'', 
Universit\`a di Salerno, Via Ponte Don Melillo, 84084 Fisciano (SA), Italy}
\affil{$^2$ INFN, sezione di Napoli, Italy}
\affil{$^3$ Istituto Internazionale 
per gli Alti Studi Scientifici (IIASS), Vietri Sul Mare (SA), Italy}
\affil{$^4$ INAF-OAC, Italy}
\affil{$^5$ Department of Astronomy, 
Ohio State University, 140 West 18th Avenue, Columbus, OH 43210, US}
\affil{$^6$ INAF-OAB, Italy}
\affil{$^7$ Dipartimento di Fisica, Universit\`a del Salento and INFN, Sezione
di Lecce, CP 193, 73100 Lecce, Italy}
\affil{$^8$ SUPA, University of St Andrews, School of Physics \& Astronomy, North
Haugh, St Andrews, KY16 9SS, United Kingdom}
\affil{$^9$ Institute for Theoretical Physics,  University of Z\"urich, 
Winterthurerstrasse 190, 8057 Z\"urich, Switzerland}
\affil{$^{10}$ Dipartimento di Ingegneria, Universit\`a del Sannio,
Corso Garibaldi 107, 82100 Benevento, Italy}
\affil{$^{11}$ XMM-Newton Science Operations Centre, ESAC, 
ESA, PO Box 78, 28691 Villanueva de la Ca{\~n}ada, Madrid, Spain}
\affil{$^{12}$ Dipartimento di Fisica, Politecnico di Torino, 
Corso Duca degli Abruzzi 24, 10129 Torino, Italia}
\affil{$^{13}$ INFN, Sezione di Torino, 10125, Torino, Italia}

\begin{abstract}
We present an updated analysis of the M31 pixel lensing candidate event OAB-N2 
previously reported in \cite{novati09}.
Here we take advantage of new data both astrometric and photometric.
Astrometry: using archival 4m-KPNO and HST/WFPC2 data we perform
a detailed analysis on the event source whose result, although not fully conclusive
on the source magnitude determination,
is confirmed by the following light curve photometry analysis.
Photometry: first, unpublished WeCAPP data allows us to confirm
OAB-N2, previously reported only as a viable candidate, 
as a well constrained pixel lensing event.
Second, this photometry enables a detailed analysis in the event parameter
space including the effects due to finite source size.
The combined results of these analyses allow us to put a strong
lower limit on the lens proper motion.
This outcome favors the MACHO lensing hypothesis
over self lensing for this  individual event 
and points the way toward distinguishing between
the MACHO and self-lensing hypotheses from larger data sets.
\end{abstract}


\section{Introduction} 

Gravitational microlensing is an established
tool of research. The original
motivation \citep{pacz86} has been the search for dark matter
in the form of massive compact halo objects (MACHOs)
(for a recent discussion from a larger
point of view on dark matter and gravitational lensing we refer to \citealt{massey10}). 
The first lines of sight explored to this
purpose have been those toward the Magellanic
Clouds (LMC and SMC). Microlensing campaigns
toward the Galactic center have then been used to constrain
Galactic structure (\citealt{moniez10} and references therein)
and, more recently, for the detection
of extra-solar planets (\citealt{gaudi10,dominik10} and references therein).

Results on MACHOs toward the LMC and the SMC
have been reported by
the MACHO \citep{macho00}, the EROS \citep{eros07}
and the OGLE collaboration \citep{ogle09}.
There is agreement that MACHOs are excluded
as a major component of
dark matter for a wide range of possible 
mass values (down to about $10^{-6}~\mathrm{M}_\odot$). 
There remains however a range, $(0.1-1)~\mathrm{M}_\odot$, 
in which the results of the MACHO collaboration,
as confirmed by \cite{bennett05}, indicate that MACHOs
might compose about a $f=20\%$
fraction of the Milky Way (hereafter MW) halo mass
at least up to the distance of the LMC.
This outcome is challenged by the results
of the EROS and the OGLE collaborations.
A possible contamination of the microlensing signal
due to MACHOs, once the background of intrinsic variable 
stars that can mimic microlensing is excluded, 
is \emph{self lensing}, wherein both the source and lens
belong to some luminous 
population\footnote{Hereafter we will refer
to ``self lensing'', broadly speaking, for any 
microlensing event with a \emph{stellar} lens,
not necessarily belonging to the \emph{same} stellar population
of the source.}. (This possibility has been first
considered in \citealt{sahu94,gould95} and further
studied by several authors, e.~g. \citealt{gyuk00,mancini04};
we refer also to \citealt{novati09b} for a specific 
analysis of the OGLE-II results, to \citealt{novati06} for a study
of the possible contribution of the LMC own dark matter
halo to the observed MACHO events and to \citealt{moniez10}
for an updated review on this issue). Indeed, the fact that
the only remaining allowed mass range for MACHOs,
from the LMC analyses, corresponds 
to that of the stars that might also
act as lenses may be indicative of some observational bias
(we recall that the characteristics
of the microlensing events, in particular their
duration, depend on the lens mass). 
On the other hand, if the reported events are 
really to be attributed to MACHOs, a fortiori
for such a sizeable fraction as that implied
by the MACHO collaboration results,
this might have some deeper astrophysical meaning.

Beyond the Magellanic Clouds, the next suitable
target for a microlensing search is our nearby and similar
galaxy of Andromeda, M31 \citep{crotts92,agape93,jetzer94}.
It allows one to explore a different line of 
sight through the MW halo; we can fully map
M31's own dark matter halo (which is not
possible for the MW, this being perhaps
the most severe limitation of LMC studies);
finally, the inclination of M31 is expected
to give rise to a characteristic signature
in the spatial distribution of M31 halo events,
so facilitating their identification against
the contamination of self-lensing events.
Because of the distance of M31, the sources
for microlensing events are no longer resolved objects,
at least for ground-based telescopes,
so that we enter the \emph{pixel lensing} regime \citep{gould96}. 
Several observational
campaigns have already been 
undertaken along this line of sight (for a review we refer
to \citealt{novati10}). The results obtained up to now
on the MACHO issue are controversial.
The POINT-AGAPE collaboration claimed evidence of a MACHO signal
in the same mass range indicated by the MACHO 
LMC analysis \citep{novati05}.
This outcome, however, has been challenged by
the MEGA collaboration who
concluded that the detected signal is compatible
with the expected self-lensing rate \citep{mega06}.
Finally, the WeCAPP collaboration
is expected to present the final 
analysis of their 11-year campaign\footnote{A.~Riffeser 
and S.~Seitz, private communication.}.

The issue of self lensing for M31 microlensing
is even more difficult than for the LMC
for two principal reasons. First, the expected self-lensing
rate is quite large with respect to MACHO lensing,
at least in the inner part of M31 (in which both
signals have the larger expected rate because
of the huge number of available sources). The exact ratio
depends not only on the unknown halo fraction in the form of MACHOs
and on the MACHO mass but also on the not fully
understood contribution of M31 \emph{stellar} lensing.
(In fact, a possible approach to deal with this problem is that
followed in the already cited analyses of \citealt{novati05}
and \citealt{mega06}, where statistical arguments were 
used for a full set of events. Interestingly,
a main point of disagreement between these two analyses
is the role played by self lensing with respect to MACHO lensing,
as discussed also in \citealt{ingrosso06,ingrosso07}).
Second, the lack of knowledge
of the source flux, specific for the ``pixel'' lensing regime,
adds a further degeneracy in the event
parameter space that makes  the disentangling
of the two signals more difficult. In particular, this degeneracy implies
that, rather than determining the Einstein time, $t_\mathrm{E}$,
the impact parameter, $u_0$, and the source flux $\phi^*$,
from the light curve  one can usually
reliably estimate only the full-width-half-maximum duration,
$t_{\mathrm{FWHM}}=t_\mathrm{E}\,f(u_0)$ \citep{gondolo99},
and the flux deviation at maximum amplification,
$\Delta\phi=\phi^* \, A_\mathrm{max}(u_0)$, usually
expressed in term of magnitude, e.~g. $\Delta R$ 
\citep{gould96,wozniak_pacz97}.

This problem of interpretation of the lens nature,
together with the overall small number of \emph{reliable} microlensing
candidate events reported toward M31 so far, make 
the issue of a correct astrophysical understanding of \emph{single}
events that may contain precious additional information 
particularly relevant. 
For M31 lensing this issue has been first
emphasized by the WeCAPP collaboration in \cite{arno06}. 
The WeCAPP collaboration, furthermore,
presented an extremely detailed analysis of the
microlensing event PA-S3/GL1 (first reported by POINT-AGAPE, 
\citealt{paulin03,belokurov05}, and subsequently, \citealt{wecapp03},
detected in their own data set
by the WeCAPP collaboration itself). Also thanks
to the excellent sampling along the bump assured by the
two independent data sets, and on the basis
of a study of the \emph{source finite size effect} \citep{witt_mao94}, 
\cite{arno08} showed that
this extremely bright microlensing event is more likely
to be attributed to MACHO lensing than to self lensing.
Interestingly, this would hold even though the event is located
very near to the M31 center, $d\sim 4'$, where the
expected self-lensing rate is larger.

In fact, the first detailed analysis of a M31 pixel lensing event
aimed at the characterization of the lens nature
has been presented by the POINT-AGAPE collaboration
for the event PA-N1 \citep{point01}.
With the source magnitude and color fixed
thanks to an analysis of archival \emph{Hubble Space Telescope} (HST) data,
and based on a specific Monte Carlo simulation of the experiment,
where the observed characteristics
of the event had been used to delimit the possible parameter space
(using in particular a lower limit for the lens proper motion), 
\cite{point01} conclude that the PA-N1 lens
is more likely a MACHO if the mass halo fraction 
in the form of compact halo object is above $20\%$.
The results of this analysis have been however 
challenged by the MEGA collaboration 
(who also reported this event as MEGA-ML16, \citealt{mega06}),
in particular by \cite{mega05_hst} who argued, 
on the basis of new HST observations, 
against the source identification reported in \cite{point01}.

Following the strong interest in the characterization
of single events, in the present work we will present an updated
analysis of the microlensing candidate event OAB-N2
first reported by the PLAN (Pixel Lensing 
ANdromeda)  collaboration in \cite{novati09}.
First, we present a new analysis related to the issue
of the source identification and characterization 
using KPNO and HST archival images, Sect.~\ref{sec:source}.
The driving motivation for a new analysis comes from
the significantly improved sampling along the flux deviation
due to the access to unpublished data of the WeCAPP collaboration
\citep{wecapp01}. As we show, these new data allow us
to firmly establish  the microlensing nature of this flux variation
and to carry out an improved and more detailed analysis 
of the lensing parameter space with the inclusion
of the finite source size effect. 
This allows us to put some constraints on the relative lens-source \emph{proper motion} 
\citep{gould94}, Sect.~\ref{sec:light}.
In Sect.~\ref{sec:res} we discuss the outcomes of the analysis 
within the framework of the Monte Carlo 
simulation of our M31 pixel lensing experiment.
The limits on the lens proper motion are here relevant
as they can be used as a tool to distinguish 
bewteen the self lensing and MACHO lensing nature
of this microlensing event.

\section{Analysis} \label{sec:ana}

\subsection{Source identification and characterization} \label{sec:source}

In \cite{novati09} we have tentatively identified the OAB-N2
candidate event source with a rather bright star, $R=20.9$ $R-I=1.2$,
the nearest \cite{massey06} catalog object to our estimated 
event position. Also motivated by the updated light curve analysis
that suggests a fainter source star (Sect.~\ref{sec:light})
we have performed a more detailed analysis. As a result we find
this catalog star \emph{not to be} the event source. However,
the available data do not allow us to conclusively
identify and characterize the candidate event source.

More specifically, we have proceeded as follows.
We have downloaded the images used by \cite{massey06}
for their analysis, collected at the 4m Mayall-KPNO 
telescope\footnote{http://archive.noao.edu.}.
In particular we have used the $I$ band image M31F5I-1.fits
with exposure time of $150~\mathrm{sec}$.
We have selected a sample of $\sim~200$ stars brighter than $R_C=20$
identified both on M31F5I-1 and our OAB image to carry out a pixel-to-pixel 
relative astrometry calibration\footnote{The KPNO and OAB images 
have pixel scales of $0.26''$ and $0.58''$, respectively.}, 
for which we get to $\mathrm{rms}\sim 0.2''$.
Because at maximum amplification the event is a resolved object,
the accuracy of the event position determination
is better, below $0.1''$.
Taking therefore $\sigma=0.2''$ we find that
the catalog star previously identified 
as the event source is in fact about $4.3\sigma$ away from
the event position (and this is also confirmed 
to be the star reported in the catalog
nearest to the event position), whereas a fainter object,
not present in the catalog but still clearly visible
(as checked also on a few other \citealt{massey06} images),
is located at only $1.3\sigma$ (hereafter we refer to the catalog
object as S1 and to the fainter one as S2).
This result is confirmed by an independent analysis in which both the KPNO and the OAB
images are first ``astrometrised'' with respect to the USNO-B1 catalog
\citep{monet03}. This analysis, however, has worse
precision since the nearest catalog objects that avoids
the inner M31 central region are located at about $3'-4'$
from the event position. 

As a second step we perform PSF photometry analysis
of the KPNO $I$-band image. For S2 we estimate $I_C=20.5\pm 0.2$. 
As expected, this is fainter than S1,
by about $\sim~1~\mathrm{mag}$, but the statistical
error turns out to be extremely large, more than twice greater than 
the average error we find for stars of similar
brightness in this image. Furthermore, we find also the output fit \emph{sharpness}
parameter for S2 to be a clear outlier. Both these elements
indicate that S2 is more likely to be a \emph{blend} of two or more stars.

To investigate further this issue we have looked for 
archival HST images of this field. 
We find  2~F606W and 3~F814W WFPC2/WF3  images (pixel scale $0.1''$) 
available  (GO 5971, P.I. Griffiths), out of which, however,
only a $1000~\mathrm{sec}$ exposure time F814W image
is reported not to have quality problems, which we then use
for our analysis. The cross-match
of the KPNO image with the HST one is carried out by 
the identification of 12 stars on both images and we end up 
with a relative astrometry precision of $\mathrm{rms}\sim 0.1''$.
Corresponding to the KPNO S2 position we find
two stars (as identified also in the HST photometry analysis)
lying within a distance of about $0.3''$ (corresponding roughly
to one KPNO pixel). This confirms our previous conclusion
on S2 being a blend.
Both of these stars lie within $3\sigma$
of the estimated event position, with
the brighter one lying nearer (about $1\sigma$).
Finally, we have performed a photometry analysis on the HST image.
In particular we have made use of both  
the HST Stetson\footnote{P.~Stetson, private
communication.} PSF and of the HSTphot WFPC2 
photometry package \citep{dolphin00}, for which
we have considered three options for the background fit and PSF evaluation.
In accord with this latter analysis we find, for the brighter
of the two sources lying at the S2 position, 
$I_\mathrm{instr}\sim 20.6-21.1$\footnote{We recall that
the F814W filter is similar to the standard Cousin-$I$.},
depending on the photometry option, with a formal statistical error
$\sigma\sim~0.1~\mathrm{mag}$,
and a magnitude difference between the two sources of $\sim~0.2~\mathrm{mag}$.
For the first analysis we find a somewhat brighter magnitude value
for the brighter object, although with larger error,  $\sigma\sim~0.2-0.3~\mathrm{mag}$, 
and larger  magnitude difference, $\sim~0.8~\mathrm{mag}$.
Taking the HSTphot analysis as the fiducial one,
we get to a final estimate of $I_\mathrm{instr}=20.8\pm 0.3$
for the brighter star.
The lack of color information,
the large M31 background level and crowdedness at the event position
which makes the photometry less reliable,
and the presence of a second, fainter, star 
as a viable source  do not allow us to consider 
these results on the HST photometry as conclusive.

Overall, this new astrometry and \emph{image}-based photometry analysis,
although with the already stressed caveats, allows us to 
definitively affirm that the OAB-N2 source is a fainter star
than previously concluded. This result turns
out to be in agreement with the outcome
of the new photometry \emph{light curve}-based analysis that
we will detail in the following Section.

\subsection{Light curve analysis} \label{sec:light}

\begin{figure}
\epsscale{1.}
\plotone{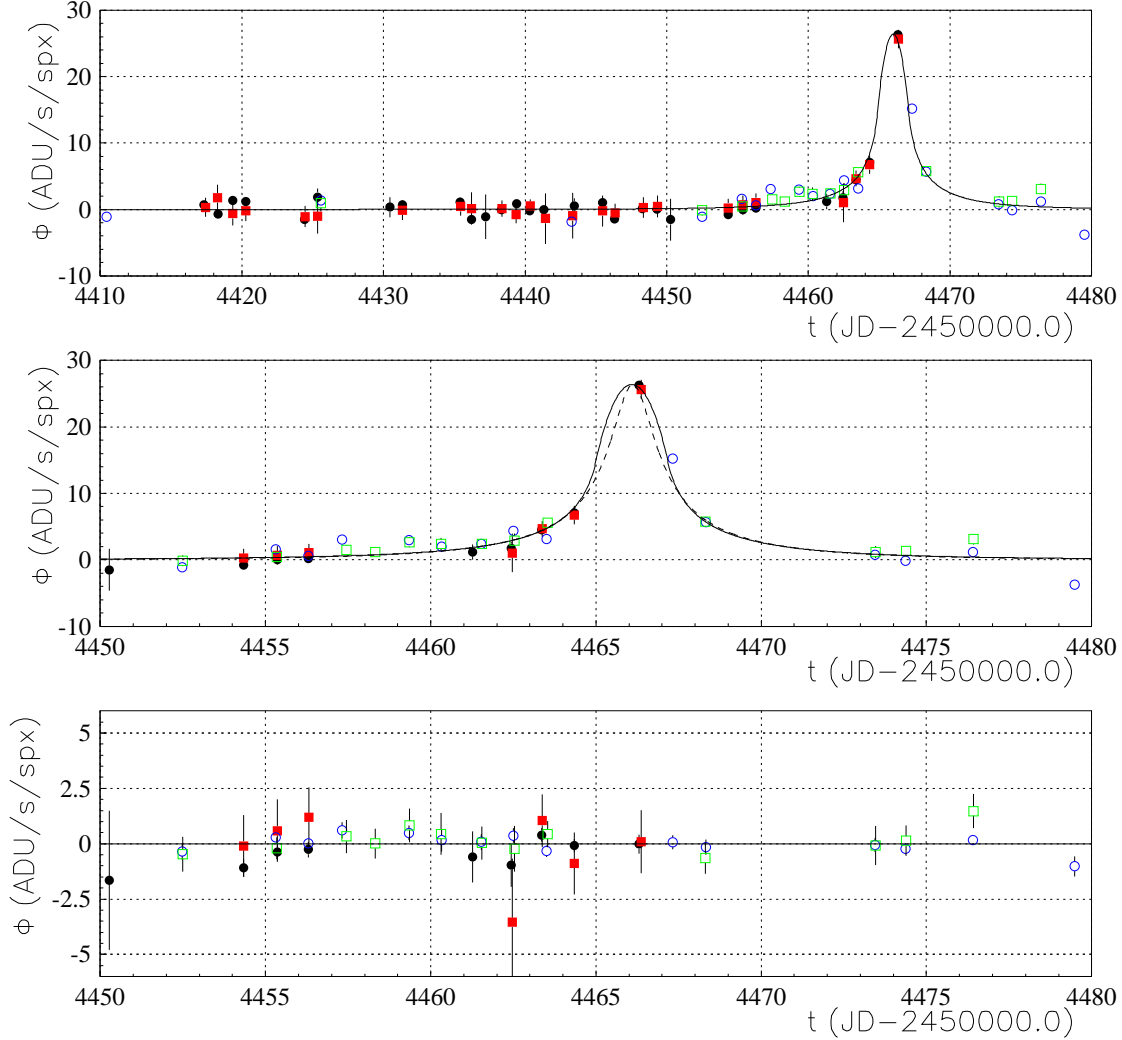}
\caption{
OAB-N2 light curve  and residuals
with respect to the best fit. 
Circles (boxes) are for $R$ ($I$)-band,
with filled (empty) symbols for OAB (WeCAPP) data.
Top and middle panels:
the solid (dashed) lines show the best fit with (without) inclusion of
finite size source effect.
The abscissa units are time in days (JD-2450000.0).
The ordinate units are flux in ADU s$^{-1}$ per superpixel.  
\label{fig:cl}
}
\end{figure}

The microlensing flux variation OAB-N2 has
been first reported in \cite{novati09}
in the framework of a long-term pixel lensing monitoring
of M31 by the PLAN collaboration at the 152cm Loiano telescope \citep{novati07}.
This appears as a bright, $\Delta R\sim 19$, short duration,
$t_{\mathrm{FWHM}}\sim 3~\mathrm{days}$,
flux variation, lying at about $3'$ from
the M31 center, with a flat baseline
on both previous OAB and POINT-AGAPE data sets.
These characteristics,
the excellent agreement with a \pacz\ shape
together with the rather red color,
$R-I\sim 1.1$ (this feature is against the alternative
hypothesis of contamination from
some kind of eruptive variable,
usually bluer), have all been considered 
as strong indications of the genuine
microlensing origin of this flux variation.
However, the analysis presented in \cite{novati09}
suffered from the serious limitation in that
data points were available on the rising part
of the flux variation only. This 
did not allow us to test the expected symmetric
shape characteristic of microlensing events
and to definitively conclude on the microlensing
nature of this flux variation.
We are now able to overcome this problem
as data points along the descent of the bump
have been made available
to us from the WeCAPP collaboration \citep{wecapp01}. 
These data have been collected at the $0.8~\mathrm{m}$
Wendelstein telescope in two pass bands, $R$ and $I$,
similar to those used for OAB data.

\begin{center}
\begin{deluxetable}{rrrrrrrrrr}
\tablecolumns{10}
\tablewidth{0pc}
\tablecaption{OAB-N2 : characteristics and \pacz\ fit results (corrected for finite source size
and limb darkening effects) \label{tab:res}}
\tablehead{
\colhead{$\alpha$} & \colhead{$\delta$} & \colhead{$\chi^2$} & 
\colhead{$t_0$ (JD-2450000.0)} & \colhead{$u_0$} &
\colhead{$\rho$} & \colhead{$t_\mathrm{E}$ (days)} & 
\colhead{$t_*$} & \colhead{$R_*$} & \colhead{$R-I$}
}
\startdata
$0^\mathrm{h} 42^\mathrm{m} 50^\mathrm{s}.31$ & $41^\circ 18' 40''.1$ & 326.16 
& $4466.23^{+0.13}_{-0.10}$ &
$0.07^{+0.07}$ & $0.154^{+0.062}_{-0.043}$ & $8.1^{+2.7}_{-2.1}$ &
$1.245^{+0.10}_{-0.13}$ & $22.25\pm 0.51$ & $1.24\pm 0.03$\\
\hline
\enddata
\tablecomments{It results $\chi^2/\mathrm{dof}=1.24$. The reported
error are for $\Delta\chi^2=1.0$ (68.3\% level). For the impact
parameter $u_0$ there is no minimum value at the requested
confidence level (as shown also in Fig.~\ref{fig:chi2}). 
For the (normalized) angular radius $\rho$ we find
also a secondary minimum at $\rho\sim 0.08$ 
(correspondingly  $u_0\sim 0.11$ and $t_*\sim 0.55$).
The source magnitude and color values are \emph{not} corrected for extinction.}
\end{deluxetable}
\end{center}

The analysis of the joint OAB-WeCAPP light curve, shown
in Fig.~\ref{fig:cl}
together with the results of the fit to be discussed below,
is extremely interesting.
WeCAPP data alone, missing the peak of the flux variation,
do not allow to detect this flux variation. They are however
essential to constrain the flux variation as they nicely
cover the descent and contribute to
the already excellent coverage along the rising
wing of the flux variation, extremely
important to robustly constrain the microlensing
parameters \citep{baltz00,dominik09a}. The excellent agreement with
a \pacz\ shape and the achromaticity are now probed
along all of the flux variation.
\emph{We confirm the genuine microlensing nature of OAB-N2}. 
This conclusion is enhanced by the fact
that we are considering two completely independent data sets
analyzed, moreover, following two different schemes of photometry reduction,
the ``superpixel'' photometry with empirical
correction of seeing variations for OAB data \citep{agape97,novati02}
and difference image analysis for WeCAPP data \citep{alard_lupton98,gossl02}.
For a recent analysis on difference imaging photometry
specific for high surface brightness targets  such as the M31 bulge
region we refer to \cite{kerins10}.

The excellent sampling along the bump allows us to
move further in the analysis and to look
for a better characterization of this microlensing event.
In particular we want to look for signatures of \emph{finite source} effect
\citep{witt_mao94}
with the purpose to eventually constrain 
the lens \emph{proper motion} \citep{gould94}.

The relevant parameter for the finite source effect
is $\rho\equiv \theta_*/\theta_\mathrm{E}$, where $\theta_*$
is the angular radius of the source and $\theta_\mathrm{E}$
the angular Einstein radius. We recall that finite source
effects become relevant whenever $\rho$ gets larger
than the impact parameter $u$.
For the event amplification, following the scheme outlined 
in \cite{yoo04}, we write
\begin{equation} \label{eq:amp}
A\left(t\right) = - A_\mathrm{pacz}\left(t\right) 
B_1\left(z\right) \Gamma_\lambda + A_\mathrm{fs}\,.
\end{equation}
Here $A_\mathrm{pacz}$ is the standard \pacz\ amplification,
$A_\mathrm{fs}$ is the amplification including the finite source effect,
for which we use the exact \cite{witt_mao94} expression 
(for a different approach we refer to the recent analysis in \citealt{lee09}),
$z\equiv u/\rho$, and $\Gamma_\lambda$ are the, wavelength dependent, linear 
limb darkening coefficients. 
The expression for $B_1(z)$ is given in Eq.~16 of \cite{yoo04}.

The inclusion of limb darkening is relevant as we know
the source flux not to be uniformly distributed across the star surface.
To evaluate the $\Gamma_\lambda$ we make use of the results of \cite{vanhamme93}.
Starting from the event color (the driving
and essential parameter) we have to further specify 
the star temperature, surface gravity and metalicity\footnote{We make use of the
relationship $\Gamma = 2u/(3-u)$ to link $\Gamma$ to the linear limb
darkening parameter $u$ (not to be confused
with the microlensing impact parameter) given in \cite{vanhamme93}.}. 
To this purpose we make use of the \cite{marigo08} isochrones.
We consider two cases: a $12~\mathrm{Gyr}$ population with
$Z=0.030$, typical for the bulge, and a $2~\mathrm{Gyr}$
population with $Z=Z_\odot=0.019$, typical for the disk \citep{arno06}.
From the light curve analysis, and the magnitude
zero point, we estimate a dereddened color $(R-I)_C=1.24$.
Taking into account the MW foreground extinction, 
$E(B-V)=0.062$ \citep{schlegel98},
we get $(R-I)_{0,C}=1.20$.
For bulge/disc sources we then get $T_\mathrm{eff}\sim 3600/3550~\mathrm{K}$,
$\log(g)\sim 0.90/0.60$. Finally we get $\Gamma =0.81/0.81\,(0.62/0.60)$
for $R_C$ ($I_C$), respectively. Even if we include
a term of internal extinction \citep{arno06} the final result does not change
significantly. Using $\mathrm{ext}_R=0.19$ for the bulge 
we finally get to $\Gamma =0.80\,(0.59)$, and with
$E(B-V)=0.22$ for the disc we get to evaluate $\Gamma =0.76\,(0.58)$.
In conclusion, we use $\Gamma_\lambda=0.80, 0.60$ for $R$ and $I$
band data. We reach consistent results basing our analysis on 
the \cite{claret00} limb darkening coefficients instead.

We carry out the light curve \pacz\ fit (corrected for finite source size
and limb darkening effects) using Eq.~\ref{eq:amp} for the
microlensing  amplification\footnote{We rescale the error bars, by $\sim 10\%$,
so to force, along the baseline, $\chi^2/\mathrm{dof}$ to unity.
This only marginally, however, changes the outcome of the analysis.}.
For a simultaneous fit of the two joint data sets
with two bands each,
overall we have 12 independent parameters
(4 and 8 to account for geometry and fluxes, respectively).
The microlensing amplification parameters: the Einstein time, $t_\mathrm{E}$,
the minimum value of the impact parameter, $u_0$, the (normalized)
source angular radius, $\rho$ and the time at maximum amplification, $t_0$. 
The 8 flux parameters are, for both $R$ and $I$ 
and OAB and WeCAPP data,
the background value and the source flux,  $\phi_R$ and $\phi_I$.

Our first purpose is to constrain the 
normalized source angular radius, $\rho$,
or equivalently $t_*\equiv \rho\,t_{\rm E}$,
the source radius crossing time
(the Einstein time turns out to be better
constrained with respect to $\rho$).
Then, since the event source flux and color can be used to estimate 
the source angular radius, $\theta_*$,
to finally constrain $\mu$, the lens proper motion
\begin{equation} \label{eq:mu}
\mu = \frac{\theta_\mathrm{E}}{t_\mathrm{E}}=\frac{\theta_*}{t_*}\,.
\end{equation}

\begin{figure}
\epsscale{1.}
\plotone{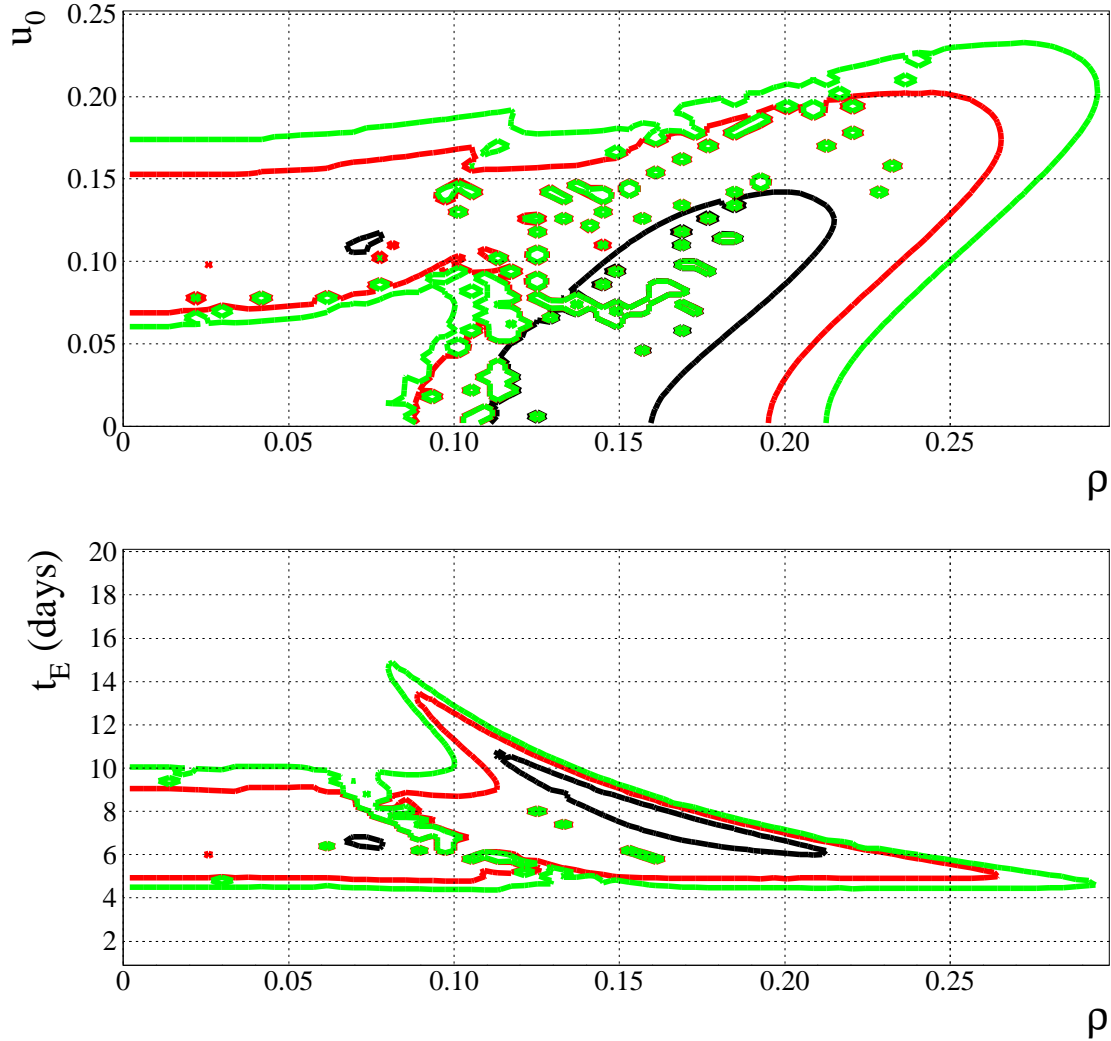}
\caption{
68.3\%, 90\% and 95\% $\chi^2$ contour maps in the parameter space $\rho$-$u_0$
(top panel) and $\rho$-$t_\mathrm{E}$. $\rho$, $u_0$ and $t_\mathrm{E}$
are the normalized angular radius of the source, the impact parameter
and the Einstein time, respectively.
\label{fig:chi2}
}
\end{figure}

\begin{figure}
\epsscale{1.}
\plotone{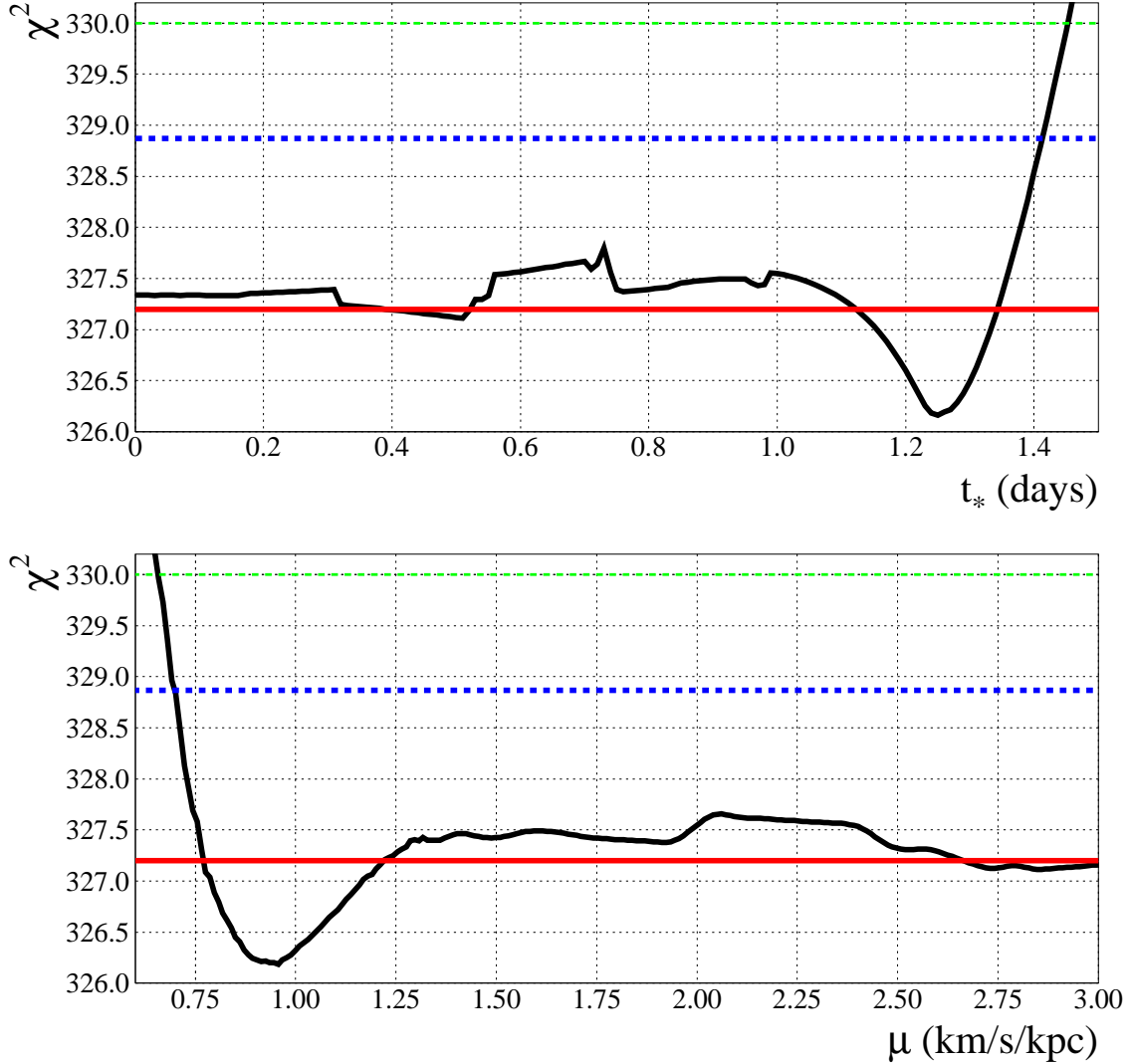}
\caption{
Plots of $\chi^2$ vs the source radius crossing time ($t_*$, top panel) and $\chi^2$ vs 
the lens proper motion ($\mu$).
Bottom panel: we use an estimate of the source angular radius 
of $0.67~\mu\mathrm{as}$ (Sect.~\ref{sec:light}).
The solid, dashed and thin dashed lines indicate the 68.3\%, 90\% and 95\%
confidence level, respectively.
\label{fig:chisq_ts_mu}
}
\end{figure}

As a tool we make use of a $\chi^2$ analysis
in which we scan the parameter space of one (or more) fit parameters 
while  the remaining are left free and determined, at each step of the scan, 
by the fit\footnote{We use, within the CERNLIB MINUIT package, 
the minimization tool 
MIGRAD  ``a variable-metric method with inexact line search, 
a stable metric updating scheme, and checks for positive-definiteness''
http://wwwasdoc.web.cern.ch/wwwasdoc/minuit/minmain.html.}.
The main results of this $\chi^2$-based analysis, discussed below,
are presented in Table~\ref{tab:res} and Fig.~\ref{fig:chi2},\ref{fig:chisq_ts_mu}.

A first relevant outcome is the excellent agreement
with a \pacz\ shape reflected in the small value $\chi^2/\mathrm{dof}=1.24$.
The best fit value for the source magnitude is 
significantly \emph{fainter} than in our previous \cite{novati09}
analysis, although affected
by a very large error.
In particular, the result is $1.3~\mathrm{mag}$ fainter (beyond 95\% level)
than the \cite{massey06} catalog source.
Furthermore, it is in agreement with
the HST-based estimate in Sect.~\ref{sec:source}
(for $I$-band magnitude $21.0\pm 0.5$ to be compared with $20.8\pm 0.3$).
Both the source magnitude and the Einstein time  
are rather well constrained within the fit:
at 95\% level we obtain $R_*=22.2^{+0.9}_{-1.1}$
and $t_\mathrm{E}=8.2^{+6.9}_{-3.9}$, respectively.
This is not the case, however, for the normalized angular radius $\rho$,
which is degenerate with the impact parameter $u_0$
(the impact parameter is in fact unbounded from below even at the 68.3\% level) 
along two different directions (Fig.~\ref{fig:chi2}, top panel).
Specifically, we find an absolute minimum 
at $\rho\approx 0.15$ with $u_0\approx 0.07$
(along the corresponding direction of degeneracy,
$u_0$ goes to zero with $\rho>0.1$)
and then a secondary minimum, with $\Delta\chi^2\sim 0.9$, 
with $\rho\approx 0.08$ and $u_0\approx 0.11$
(here, just above the 68.3\% threshold for $\Delta\chi^2=1.0$,
$\rho$ goes to zero with roughly constant $u_0$).
The former rather large value of $\rho$, 
with $\rho/u_0\approx 2$ might be taken as an indication
of finite size source  effect. Given the caveats
further addressed below this would make
of OAB-N2 the first microlensing event reported toward M31
with a signature of finite size source effect.
Indeed, the secondary minimum,
as well as the full degenerate parameter space region for $\rho<0.1$,
just above the $\Delta\chi^2=1.0$ threshold, do not allow us to 
draw firm conclusions on this issue. 
Overall, the indication for $\rho>0$
remains extremely tenuous. In particular the 
$\chi^2$ improvement with respect to the case with no 
inclusion of the finite size parameter, $\rho$,
is only marginal, $\Delta\chi^2\sim 1$.
The inclusion of $\rho$, however, is essential
for the analysis as a tool to allows us to constrain
the lens proper motion.
The degeneracy for $\rho$ is shown also 
in the parameter space $\rho$-$t_\mathrm{E}$ (Fig.~\ref{fig:chi2}, bottom panel).
This result is reflected in the estimate of the source radius crossing
time for which we find  the absolute minimum at $t_*\approx 1.25~\mathrm{days}$,
but also a secondary minimum at $t_*\approx 0.55~\mathrm{days}$ 
(Fig.~\ref{fig:chisq_ts_mu}, top panel).

The missing information to get to the lens proper motion $\mu$
is the source angular radius, $\theta_*$. To this purpose
we use the empirical relationship $\theta_*=\theta_*(V, V-K)$
for \emph{giants} discussed in \cite{vanbelle99}.
Taking into account the MW foreground extinction \citep{schlegel98},
and the \cite{marigo08} isochrones as above
we estimate, both for a bulge or a disc source, $V=23.3$ and $V-K=5.0$.
This gives  
\begin{equation} \label{eq:thetas}
\theta_* = 0.67^{+0.17}_{-0.14} \quad \mu\mathrm{as}\,,
\end{equation}
where the error budget is dominated by the statistical uncertainty
of the source magnitude within the light curve fit. The result
we get to for $\theta_*$ is very similar, within 1\%, if we additionally
take into account the internal M31 bulge or disc extinction as discussed above.
In these cases we would find a \emph{fainter} source with a \emph{bluer} color,
and these two effects balance one each other in the evaluation
of the angular radius. (An alternative, more recent and more
accurate $\theta_*=\theta_*(V, V-K)$
relationship is discussed in \citealt{kervella04}. The result
we would get to in this case for $\theta_*$ is about 20\% larger,
still comfortably within $1\sigma$.
For our discussion we however prefer that discussed 
in \citealt{vanbelle99}
as it is derived from a sample of data with $2.0\le V-K \le 8.0$
whereas \citealt{kervella04} use a more Cepheid-specific sample
with $1.0\le V-K \le 2.4$, with the former being more suitable
for us given the source color). For the discussion
of our results on $\mu$ we finally recall the useful relationship
\begin{equation} \label{eq:thetas2}
\theta_* = a \sqrt{\phi}\,,
\end{equation}
where the coefficient $a$ depends only on the color
($\phi$ is the source flux).
In particular this allows us to carry out
the $\chi^2$ analysis (Fig.~\ref{fig:chisq_ts_mu}, bottom panel),
in the 3-d parameter space $t_\mathrm{E},\,\rho$ and $\phi$.

The value for the lens proper motion we obtain, corresponding to the best fit
value, is $\mu=0.93\,\mathrm{km}\,\mathrm{s}^{-1}\,\mathrm{kpc}^{-1}$.
The discussed large residual degeneracy for the angular radius $\rho$,
and the related aspects discussed above, lead us however to report,
as a more robust outcome of the present analysis,
an \emph{upper} limit for $\rho$ (and $t_*$) and a \emph{lower} limit
for $\mu$ only. At 68.3\% level we get
\begin{eqnarray}
\rho & < & 0.22\,,\nonumber\\
t_* & < & 1.35\quad\mathrm{days}\,,\nonumber\\
\mu & > & 0.75\quad\mathrm{km}\,\mathrm{s}^{-1}\,\mathrm{kpc}^{-1}\,.\nonumber
\end{eqnarray}
For the lens proper motion we also get 
$\mu>0.68\,(0.64)~\mathrm{km}\,\mathrm{s}^{-1}\,\mathrm{kpc}^{-1}$
at the 90\% (95\%) level, respectively.

\section{Discussion} \label{sec:res}

\begin{figure}
\epsscale{1.}
\plotone{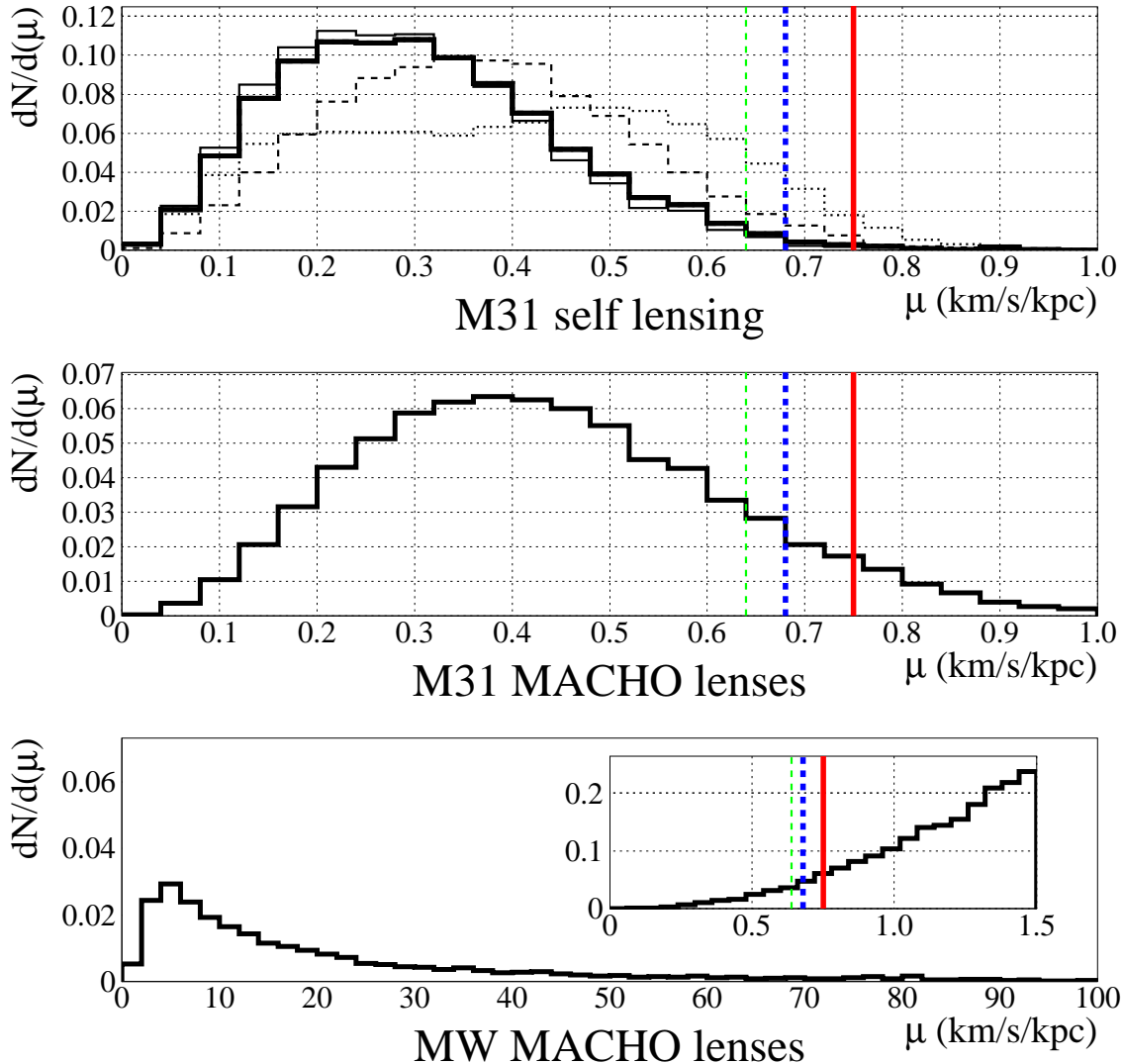}
\caption{
Expected distributions for the lens proper motion, $\mu$,
along the line of sight toward the OAB N2 position
for different source-lens populations evaluated within the Monte Carlo
simulation of the experiment.
Top panel: thin solid, dashed and dotted lines are for bulge sources
and bulge and disc lenses and disc-disc events, respectively; 
thick solid line for the overall M31 self lensing (each distribution
is normalized to 1 separately). Bulge-bulge events are expected
to contribute for about 80\% of the expected signal; MW disc
events (not shown, for which the lens proper motion distribution is expected to peak
well beyond $1~\mathrm{km}\,\mathrm{s}^{-1}\,\mathrm{kpc}^{-1}$) 
are expected to be negligibile (see text for details).
Middle (bottom) panels: M31 (MW) MACHO lenses with
the overall (M31+MW) MACHO distributions normalized to 1,
and 75\% signal expected from M31 lenses
(in the inset, bottom panel, showing a zoom for the
MW MACHO distribution, the units are $10^{-3}$).
The solid, dashed and thin dashed vertical lines (68.3\%, 90\% and 95\% level,
respectively) indicate
the \emph{lower} limit for the lens proper motion derived
in the light curve analysis (Sect.~\ref{sec:light}, Fig.~\ref{fig:chisq_ts_mu}).
\label{fig:mc_mu}
}
\end{figure}

The results of the light curve analysis presented
in Sect.~\ref{sec:light}, that in particular
confirm the outcome on the source magnitude
discussed in Sect.~\ref{sec:source},
can be used to establish the lens nature
of the OAB-N2 event. As previously discussed, the relevant issue
is to distinguish between self lensing and MACHO lensing.
To this end we focus on the lens-source relative proper motion, $\mu$,
because it is independent of the lens mass and so depends
only on the kinematics and spatial distribution of the lens population.
The potential role played by the lens proper motion,
also in relation to M31 pixel lensing, has been first
discussed in \cite{gould94,hangould96}. 
In these analyses the emphasis was given in particular
to the possibility to constrain MW MACHO  
lenses with respect to M31 lenses. This 
is relevant as MW MACHOs are expected 
to contribute, for given MACHO mass and mass halo fraction,
for about one third of the overall expected MACHO signal.
The identification of MW lenses would therefore give
a strong indication for a MACHO signal.
In the following we will point out
how and to which extent an analysis of the 
lens proper motion may allow one to distinguish
also between the different M31 lens populations.

As a tool of analysis, to compare
with the outcome of the light curve fit,
we will make use of the Monte Carlo simulation 
of the experiment described in \cite{novati09}.
As a prior we consider the line of sight toward the OAB-N2 position
(this gives however only marginal quantitative
differences from the more general case in which
the overall PLAN field of view is considered).
We also recall that as an output we have only
light curves that are \emph{selected} within the Monte Carlo 
(where in particular the experimental set up is carefully reproduced
together with a basic selection pipeline).

At the OAB-N2 position, only $\approx 3'$ from
the M31 center, the dominant contribution
from M31 self lensing is expected from events with 
bulge sources and bulge lenses. Specifically,
we expect $\sim 80\%$ of events to belong to
these populations, with the rest equally divided
between bulge-disc (both source-lens and lens-source) events,
with only a marginal 1\% of disc-disc events.
In Fig.~\ref{fig:mc_mu} we present the results
of the Monte Carlo simulation for the lens 
proper motion distributions
for different source and lens populations.
The most striking outcome for our purposes is the threshold at about 
$\mu_\mathrm{th}\sim 1~\mathrm{km}\,\mathrm{s}^{-1}\,\mathrm{kpc}^{-1}$,
such that for $\mu>\mu_\mathrm{th}$ we expect
almost no M31 lens events, while on the other hand almost
all of MW lens events are expected to fall in this region.
This is easily explained on the basis that finite size effect
is never important for MW events (for which on average
the Einstein angular radius, proportional to the square
root of the lens-source distance over the lens distance, 
is about one hundred times
larger than for M31 events 
and $\rho\propto\theta_\mathrm{E}^{-1}$) and for which, on the other hand,
the lens distance is by far smaller.
The relatively large lower limit value we find for the lens proper
motion might therefore be taken as an indication
for the lens to belong to the MW halo.
(This outcome, however, would be clearly at odd 
with the discussed marginal indication we find for
finite size source effect.)
However large, the lower limit for the lens proper motion is still
fully compatible with M31 lensing. In particular,
among M31 lensing events, we find that a broader distribution
for $\mu$ is expected for MACHO lensing,
while the limit we find is hardly
compatible with the expected distribution for bulge-bulge events.
Although we can not certainly rule out, 
for this specific event, the self-lensing hypothesis,
this outcome strongly suggests the MACHO lensing
hypothesis for the OAB-N2 event.

More specifically, corresponding to the 68.3\% lower threshold
obtained above, $\mu > 0.75~\mathrm{km}\,\mathrm{s}^{-1}\,\mathrm{kpc}^{-1}$,
for M31 MACHO lensing we still have about (for bulge source events
and similarly for disc source events that represent, however,
only about 15\% of the expected M31 MACHO signal) $6\%$ of
the expected signal (10\% and 14\% respectively at the 90\% and 95\%
levels). On the other hand, at the 68.3\% level,
we expect less than 1\% of the overall bulge-bulge signal
(still about 70\% of the overall M31 self lensing signal)
and about 1\% of bulge-disc and disc-bulge lenses.
At 95\% level these ratio rise to almost 2\% for bulge-bulge
events (about 60\% of the overall M31 self lensing signal)
and to almost 5\% for bulge-disc and disc-bulge lenses.
Although in an absolute sense M31 disc-disc events 
give only a marginal contribution, it is worth noticing
that, along this M31 central line of sight, they 
show a marked bimodal distribution for the lens proper motion
with a long tail for large values of $\mu$.
In particular, larger than the 68.3\%, 90\% and 95\% 
threshold values we still find, respectively,
2.6\%, 7.2\% and 12\% of the expected events
belonging to this population.
Overall, at the 68.3\%, 90\% and 95\% levels, we expect respectively 
about 0.6\%, 1.3\% and 2.1\% of the M31 self-lensing signal.

Besides M31 bulge and disc lenses, in principle
one might consider also \emph{Galactic} disc lenses.
Using standard models for the MW disc \citep{hangould03,novati08}
the ratio of the expected number of events for this population 
is below 1\% with respect to M31 
self lensing\footnote{Furthermore, a sizeable
fraction of these MW events is expected, because
of the extremely small observer-lens distance, to have
too bright lenses for being suitable candidates.}.
This result, joined to the overall outcome
of one, possibly two, microlensing events
reported from the 2007 PLAN campaign \citep{novati09} 
makes the possibility for OAB-N2 to be a MW disc event
extremely unlikely. 
However, for MW lensing, the lens proper motion is much
larger than for M31 lensing and this applies even more
strongly to stellar lensing than MACHO lensing for these
two galaxies.  In fact, typical values
are well above $1~\mathrm{km}\,\mathrm{s}^{-1}\,\mathrm{kpc}^{-1}$,
larger, on average, than for MW MACHOs. Hence, in particular
almost \emph{all} of the expected MW disc events
fulfill the lower limit condition we find 
on the lens proper motion. 
Larger than the 95\% level lower limit, the even small 
fraction of expected M31 self lensing events
is however still larger than the overall expected MW disc signal. 
The overall small number of expected MW disc events makes
therefore this contribution always negligible
(so that in particular we do not show it
in Fig.~\ref{fig:mc_mu}).

As for the Einstein time scale, which is rather well constrained
from the light curve fit, comparing to the
Monte Carlo distributions we find
the fit result to match the expected self-lensing
signal characteristics well. For MACHO lensing the result
is dependent on the lens mass. The fit result
is fully compatible with MACHO in the mass range
$(10^{-2}-10^{-1})~\mathrm{M}_\odot$, but still
compatible with even larger MACHO mass as 
$0.5~\mathrm{M}_\odot$.

Finally, we comment on our result for
the finite source size effect, with the best fit value
indicating a signature for $\rho>0$. As already pointed 
out this effect is expected for M31 lenses only.
Still, the value we find (as reflected also
in the proper motion best fit value) has
an a priori low probability. Together with
the only marginal detection, with a $\Delta\chi^2\sim 1$
to separate from the case with no finite size effect,
this makes this detection only ``possible''.

\section{Conclusion}

In the present paper we have presented an updated
analysis of the M31 microlensing candidate
event OAB-N2 first
reported by the PLAN collaboration in \cite{novati09}.
In this previous work, in particular,
the insufficient sampling along the flux variation
did not allow us to probe unambiguously
the microlensing nature of this flux variation.
Here we have taken advantage of newly
available WeCAPP data to complete the light curve sampling. 
From a methodological perspective,
we observe that the importance of merging
different data sets for microlensing observation
might look even trivial when considering that Galactic Bulge 
events are routinely observed by multiple telescopes 
when searching for planetary anomalies.
The present analysis, however,
shows the extent to which this is suitable, feasible and,
as this case nicely shows, possibly essential also for M31 
pixel lensing events. (Such a strategy, besides, is unavoidable
in order to achieve the challenging purpose
of a systematic research for exoplanets in M31,
\citealt{covone00,baltz01,chung06,ingrosso09}.
A first attempt in this sense is being carried 
out by the ANGRSTOM collaboration
\citealt{kerins06,darnley07,kim07}. The usually short 
expected timescale, $t_\mathrm{FWHM}$, makes however potentially
important such a strategy also for ``normal'' single-lens events.).

A preliminary analysis on archival images of the 4m Mayall-KPNO telescope 
and of WFPC2/HST allowed us to reject the tentative
identification of the OAB-N2 source made in \cite{novati09} with a rather bright 
\cite{massey06} M31 catalog source. The event source, in fact, appears
to be a nearby fainter blended object on the KPNO image,
as confirmed also by the HST analysis. 
Although we can not conclusively determine the source magnitude,
because of blending and the huge background level,
even this partial result is a first step toward 
a better characterization of this event. Indeed,
the new light curve photometry analysis confirms
that the source must be a fainter object than
previously concluded.

The improved sampling along the OAB-N2 light curve then allows
us to reach a few important conclusions. First,
we find a strong confirmation of the microlensing nature
of this flux variation. The flux variation characteristics,
the large signal to noise ratio,
the short duration and the position,
within the inner $3'$ of M31 where the expected rate
is larger, together with the further evidence
provided by the joint analysis of two
completely independent data sets, put this flux variation
among the most convincing and interesting 
M31 microlensing events reported so far.
Second, we have carried out a more detailed
analysis in the event microlensing parameter space
with the specific purpose to constrain the lens nature,
whether due to self lensing or to MACHO lensing.
The main outcome we get to is 
a \emph{lower} limit for the relative
source-lens proper motion, $\mu$. As we have shown,
this parameter, independent of  the lens mass
and therefore less subject to model issues,
not only may allow one to distinguish
M31 lensing from MW lensing (where the main
expected signal is MACHO lensing and for which
$\mu$ is significantly larger than for M31 lensing)
but also among different M31 lens populations.
Specifically, at 95\% level, we find
$\mu  > 0.64~\mathrm{km}\,\mathrm{s}^{-1}\,\mathrm{kpc}^{-1}$.
Comparing to the output of the Monte Carlo
simulation of the experiment we conclude that,
if not to the MW halo population, this value 
is more compatible with M31 MACHO lensing rather than with
self lensing, for which only about 2\% of the overall
population is expected to fulfill this constraint.
Although the self lensing hypothesis 
can not be definitively ruled out for this single specific event
so that we are not yet in a position to strongly
discriminate self lensing from MACHO lensing,
this analysis shows the extent to which
the lens proper motion, with no other prior 
assumptions on MACHOs but their spatial distribution
and kinematics, is a powerful tool
toward this purpose.

As an intermediate step of the light curve fit $\chi^2$-based 
analysis we have looked for possible signatures
of finite size source effect (through a study
of the normalized angular radius, $\rho$) 
for which however we find only marginal indications.
We recall that an evidence in this sense
would imply that the lens belongs to some M31 populations.
The robust outcome we get to is rather an \emph{upper} limit
for $\rho$ and on the related quantity, 
the source radius crossing time, $t_*=\rho\,t_\mathrm{E}$
which, together with an estimate, from the source magnitude
and color, of the physical angular radius,
finally gives us the aforementioned \emph{lower} limit
for the lens proper motion. 

The outcome of this analysis
is  affected by the lack of an accurate and precise
estimate of the unresolved source magnitude and color, 
independent of the light curve fit,
for which a focused proposal on HST data would
be extremely welcomed.

In the framework of M31 pixel lensing,
the present analysis of OAB-N2 is the third
example, following those for the events 
PA-N1 \citep{point01} and PA-S3/GL1 \citep{arno08},
where a detailed investigation of
the lensing parameter space allowed
to draw conclusions on the lens nature. 
Interestingly, although in the present work 
with a worse confidence level with respect to that reported
in \cite{arno08} for PA-S3/GL1 and with the caveats discussed in the Introduction
as for PA-N1, in all cases the analyses
point toward the same outcome, namely that MACHO lensing
should be preferred over self lensing. 

\acknowledgments
We warmly acknowledge the WeCAPP collaboration,
in particular A.~Riffeser and S.~Seitz, 
for making available to us, prior to their publication,
the WeCAPP data relative to the OAB-N2 light curve.
SCN, VB, LM and GS acknowledge support by MIUR
through PRIN Prot. 2008NR3EBK\_002, and by
FARB of the Salerno University.

\bibliographystyle{apj}
\bibliography{biblio}

\end{document}